\documentclass[12pt,preprint]{aastex}

\newcommand{\vect}[1]{\mbox{\boldmath$#1$}}
\newcommand{\dfrac}[2]{{\displaystyle \frac{#1}{#2}}  }
\newcommand{\alfven}{Alfv\'{e}n}
\newcommand{\eqref}[1]{(\ref{#1})}

\def\lesssim{\mathrel{\hbox{\rlap{\hbox{\lower4pt\hbox{$\sim$}}}\hbox{$<$}}}}
\def\gtrsim{\mathrel{\hbox{\rlap{\hbox{\lower4pt\hbox{$\sim$}}}\hbox{$>$}}}}

\slugcomment{submitted to ApJ}

\shorttitle{Planetary Migration in Poloidal Field}
\shortauthors{Muto et al.}

\begin{document}

\title{The Effect of Poloidal Magnetic Field on Type I Planetary
Migration: Significance of Magnetic Resonance}

\author{Takayuki Muto\altaffilmark{1}, Masahiro N. Machida and
Shu-ichiro Inutsuka} 
\affil{Department of Physics, Kyoto University, \\
Kitashirakawa-oiwake-cho, Sakyo-ku, Kyoto, 606-8502, Japan}
\email{muto@tap.scphys.kyoto-u.ac.jp}

\altaffiltext{1}{JSPS Research Fellow}

\begin{abstract}
 We study the effect of poloidal magnetic field on type I planetary
 migration by linear perturbation analysis in the shearing-sheet
 approximation and the analytic results are compared with numerical
 calculations.  In contrast to the unmagnetized case, the basic equations
 that describe the wake due to the planet in the disk allow magnetic
 resonances at which density perturbation diverges.
In order to simplify the problem, we consider the case without
 magneto-rotational instability.  We perform two sets of analyses:
 two-dimensional and three-dimensional.  In two-dimensional analysis,
 we find the generalization of the torque formula previously known in
 unmagnetized case.  In three-dimensional calculations, we focus on the
 disk with very strong magnetic field and derive a new analytic formula
 for the torque exerted on the planet.   We find that
 when \alfven\ velocity is much larger than sound speed, two-dimensional
 torque is suppressed and three-dimensional modes dominate, in contrast
 to the unmagnetized case.
\end{abstract}


\section{Introduction} \label{intro}

Type I planetary migration is one of serious difficulties in the theory of
planet formation and there have been a lot of work on this topic.  For an 
unmagnetized disk, extensive linear perturbation analyses have
been performed since
Goldreich \& Tremaine's pioneering work (Goldreich \& Tremaine, 1979).
The recent result by Tanaka et al. (2002) has 
shown that the protoplanet of 5M$_\oplus$ located at 5AU and embedded in
the minimum-mass solar nebula \cite{HNN85} will migrate inward to the
central star in $8\times 10^5$ years, shorter than the
observed time scale of protoplanetary disk, 
$10^7$ years (see e.g., Haisch et al. 2001).  Their results are confirmed by 
numerical calculations (see e.g., D'Angelo et al. 2003).

Most of analytic works on type I planetary migration have focused on
unmagnetized, locally isothermal disks.  Goldreich \& Tremaine (1979)
considered the two-dimensional disk, that is, there is no structure in
the vertical direction of the disk, and derived analytic formulae
of the torque exerted at Lindblad and corotation resonances.  In order
to calculate the migration rate, it is necessary to calculate the
difference of the torque exerted on the disk inside and outside of the
planet.
This is done by Ward (1986).  Later, Artymowicz (1993) derived, without
invoking WKB approximation, a generalized formula for Lindblad
resonances including the cutoff of the torque for high
azimuthal mode number.  Three-dimensional analysis is performed by
Takeuchi \& Miyama (1998), Tanaka et al. (2002), and Zhang \& Lai
(2006).  Takeuchi \& Miyama (1998) derived torque formulae for 
some resonances and full analytic calculation was performed by Zhang \&
Lai (2006).  When the planet is embedded in the equatorial plane of a
thin disk, three-dimensional modes have been shown to be ineffective
(Tanaka et al. 2002).

In contrast to all the above studies, magnetic fields are supposed to be
present in protoplanetary disks.
Significant mass accretion onto the central star requires 
an effective mechanism for angular momentum transfer. 
At present, magneto-rotational instability, or MRI 
(Balbus \& Hawley 1991), is the most likely mechanism for 
the generation of turbulent viscosity in the disks. 
However, whether the disk is magnetically active or dead depends 
strongly on the amount of gas (and hence, dust), number density, 
and the properties (size distribution etc.) of the dust grains 
because the surfaces of the dust grains are 
very efficient sites of recombination (Gammie 1996, Sano et al. 2000).  
The detailed analysis of the ionization structure with 
the standard cosmic ray ionization rate simply predicts that 
the planet forming region (approximately between 0.1AU and 10AU 
from the central star)  
might be in the dead zone where gas and magnetic field do not couple. 
However, various effects can change this prediction: 
the growth of dust grains to larger particles decreases the total 
surface area for recombination, and hence, 
increases the ionization degree.  
Sedimentation of dust grains onto the disk midplane also increases 
the ionization degree of most of the height of the disk. 
Yet another mechanism to increase the ionization rate is 
also proposed, which possibly removes the dead zone in 
the standard solar nebular model (Inutsuka \& Sano 2005).

The property of planetary migration may be totally
different if magnetic field is important.
Nelson and Papaloizou (2004) performed numerical calculations of type I
migration in a magnetized disk and indicated that the
turbulence due to MRI would result in stochastic torque on the planet.
A simple model for the random torque was presented by Laughlin
et al. (2004).  Terquem (2003) performed the linear analysis of the
torque for two-dimensional laminar disk with 
toroidal magnetic field and indicated
that when stronger magnetic field was exerted on the disk inside the
planet's orbit than the
outside, inward migration might be halted.  
Fromang et al. (2005) further investigated this situation by numerical
calculation and the numerical
results showed good agreement with linear analysis.

In this paper, we investigate the type I planetary migration when the
disk is exerted by poloidal magnetic field, which is a complementary
analysis to Terquem (2003).  As a first step to understand the
nature of migration in a magnetized disk, we perform shearing sheet
analysis, which assumes symmetric structure inside and outside the
planet's orbit, and calculate the torque exerted on one side of the
disk.  We 
restrict ourselves to a laminar disk, the case without MRI, and derive
analytic formulae of torque exerted on some 
important resonances.  We perform a three-dimensional calculation and
our formalism is a natural extension of previous studies of unmagnetized
cases that is developed by Goldreich and Tremaine (1979) and Artymowicz
(1993).
For two-dimensional modes, we derive an analytic formula which
generalizes that of Artymowicz (1993).  For three-dimensional modes, we
employ WKB approximation and derive an analytic torque formula in a
strong field limit.
We show, for three-dimensional modes, that there is a divergence in
perturbed density at certain resonances and the torque is localized at
this point.  We describe how to treat the singularity in the wave
equation.  We show that two-dimensional modes are suppressed
by poloidal magnetic field and three-dimensional modes will dominate the
total torque.  We then compare the results of the linear analysis with
a numerical calculation, and show good agreement.  
Type I migration in a disk with strong poloidal magnetic
field may be also important in the formation of planets around neutron
stars (e.g., Bailes et al. 1991).

The plan of this paper is as follows.  In section \ref{linear}, we
describe the linear analysis.  In section \ref{simu}, we describe the
numerical calculation.  Section \ref{compare} compares the
results of linear analysis and numerical calculation.  We discuss
some possible important points that are not covered in this analysis in
section \ref{discussion}, and section \ref{summary} is for our summary.

\section{Linear Calculation of Torque} \label{linear}

In this section, we derive analytic formulae for the torque exerted on
the disk by the planet by linear analysis.  The backreaction exerted on
the planet causes orbital migration.

\subsection{Basic Equations}

 For simplicity, we consider only a local region around the planet
 using shearing
 sheet model \cite{NGG87}.  Although it gives the same
 magnitude but the opposite sign of the torque between the inner and
 outer regions of the planet so the net torque becomes zero, we
 focus on one side of the disk in this setup and simplify the problem to
 understand the effect of magnetic field.
 We assume that the temperature is constant and the self-gravity of the
 disk is negligible in this local region.
 The orbit of the protoplanet is assumed to be circular on the
 equatorial plane of the disk.
 We set up local Cartesian coordinates with origin at the protoplanet's
 position and the $x$-, $y$-, and $z$-axes are radial,
 azimuthal, and vertical direction of the disk, respectively.
 We use ideal MHD equations:
\begin{equation}
 \dfrac{\partial \rho}{\partial t} + \nabla \cdot (\rho \vect{v}) =
 0
\end{equation}
\begin{equation}
 \dfrac{\partial \vect{v}}{\partial t} + (\vect{v} \cdot \nabla)
 \vect{v} = - \dfrac{1}{\rho} \nabla P - \nabla \psi_{\rm eff} - 2
 \Omega_{\rm p} (\vect{e}_z \times \vect{v}) - \dfrac{1}{4\pi \rho}
 \vect{B} \times (\nabla \times \vect{B})
\end{equation}
\begin{equation}
\dfrac{\partial \vect{B}}{\partial t} = \nabla \times (\vect{v} \times
 \vect{B})
\end{equation}
where $\rho$, $\vect{v}$, $P$, $\psi_{\rm eff}$, $\Omega_{\rm p}$,
   $\vect{e}_z$, and $B$ are the gas density, velocity, gas
   pressure, effective potential including tidal force and the planet's
   gravitational potential, Keplerian angular velocity of the
   protoplanet, a unit vector directed to the $z$-axis, and the magnetic
   flux density, respectively.
   We adopt an isothermal equation of state, $P = c^2 \rho$, where $c$
   is sound speed. The Keplerian angular velocity of the protoplanet is
   given by
\begin{equation}
\Omega_{\rm p} = \left( \dfrac{G\, M_{\rm c} }{ r_{\rm p}^3} \right)^{1/2},
\end{equation}
where $G$, $M_{\rm c}$, and $r_{\rm p}$ are the gravitational constant,
 mass of the central star, and the distance between the protoplanet and
 the central star, respectively.
 Our calculations are normalized by unit time, $\Omega_{\rm p}^{-1}$,
 unit velocity, $c$, and unit length,  
 $h \equiv c/\Omega_{\rm P}$.
 The effective potential $\psi_{\rm eff}$ in our normalization is given
 by, assuming a Keplerian disk,
\begin{equation}
\tilde{\psi}_{\rm eff} = - \dfrac{3}{2} \tilde{x}^2 \, - \, \dfrac{3
 \tilde{r}_{\rm H}^3}{\tilde{r}},
 \label{effective}
\end{equation}
 where all the quantities with tilde indicate the normalized value.
 The first term of the right hand side of equation \eqref{effective} is
 composed of the gravitational potential of the 
 central star and the centrifugal potential, and higher orders in $x$,
 $y$, and $z$ are neglected.  We also neglect the $z$-dependence of the
 gravitational potential of the central star for simplicity, and
 consider later the constant
 background density.  This greatly simplifies the calculation, and we
 have found that it does not seriously affect the results.
 The second term of the right hand side of equation \eqref{effective} is
 the gravitational potential of the protoplanet,
 where $\tilde{r}_{\rm H}$ and $\tilde{r}$ are normalized Hill radius
 and the distance from the center of the protoplanet respectively.
 The Hill radius is defined by 
$ r_{\rm H} = (M_{\rm p}/3M_{\rm c})^{1/3} r_{\rm p}$, 
 where $M_{\rm p}$ is the mass of the protoplanet.

 The background disk is assumed to have no planet.
The background gas flow has a Keplerian shear,
$\vect{v_0} = -(3x/2)\vect{e}_y$, background density is assumed to be
constant, $\rho_0$, and the background magnetic field is assumed to be
constant and poloidal, $\vect{B}_0 = B_0 \vect{e}_z$.  We denote all the
background quantities with subscript zero.

We treat the planet as a perturber on this background disk and derive
the stationary pattern excited by the planet,
 as in Goldreich \& Tremaine (1979).  
We denote perturbed quantities with $\delta$, e.g., density
perturbation is denoted as $\delta \rho$.  We Fourier transform in
$t$-, $y$-, and $z$-directions, i.e., we shall consider the
solution of the form 
$\delta \rho \propto \exp [-i(\omega t - k_y y - k_z z) ]$.
Since we consider the stationary pattern, the frequency $\omega$ is
zero.  The perturbed quantities are then
\begin{equation}
 \delta \rho(x,y,z) = \sum_{k_y,k_z} \delta \rho_{k_y,k_z}(x) e^{i(k_y y
  + k_z z)},
\end{equation}
and the inverse transformation is
\begin{equation}
 \delta \rho_{k_y,k_z} (x) = \dfrac{1}{L_y L_z} \int_{-L_z/2}^{L_z/2}
  \int_{-L_y/2}^{L_y/2} dy dz \delta \rho (x,y,z) e^{-i(k_y y + k_z z)} ,
\end{equation}
where $L_y$ and $L_z$ denote the box sizes of $y$- and $z$-directions
respectively.  Imposing periodic boundary
conditions in $y$- and $z$- directions, the wave numbers in these
directions are $k_y=2\pi n_y/L_y$ and $k_z=2\pi n_z/L_z$ respectively,
where $n_y$ and $n_z$ are integer.
We shall drop the subscripts $k_y$ and $k_z$ of the Fourier modes unless
it is ambiguous.

We define the Lagrangian displacement $\vect{\xi}$ by
\begin{eqnarray}
& \delta v_x = -i\sigma(x) \xi_x, \\
& \delta v_y = -i\sigma(x) \xi_y + \dfrac{3}{2} \Omega_p \xi_x, \\
& \delta v_z = -i\sigma(x) \xi_z,
\end{eqnarray}
where
\begin{equation}
 \sigma(x) \equiv \omega + \dfrac{3}{2} \Omega_p k_y x .
\end{equation}

Using the Lagrangian displacement, the linearized induction equations
are
\begin{eqnarray}
& \delta B_x = ik_z B_0 \xi_x, \\
& \delta B_y = ik_z B_0 \xi_y, \\
& \delta B_z = - B_0 \left( \dfrac{d \xi_x}{dx} + ik_y \xi_y \right).
\end{eqnarray}
The equation of continuity is
\begin{equation}
 \frac{\delta \rho}{\rho_0} + \dfrac{d\xi_x}{dx} + ik_y \xi_y + ik_z
  \xi_z = 0 ,
  \label{pert_EoC}
\end{equation}
and the equations of motion become, using the induction equations,
\begin{eqnarray}
 \label{pert_EoMx}
& (-\sigma^2 -3\Omega_p^2) \xi_x +2i\Omega_p \sigma \xi_y
 = -(c^2+v_A^2) \dfrac{d}{dx}
 \dfrac{\delta \rho}{\rho_0} + v_A^2 \left( -k_z^2 \xi_x - ik_z
				      \dfrac{d\xi_z}{dx} \right) -
 \dfrac{d\psi_p}{dx}, \\
 \label{pert_EoMy}
& -\sigma^2 \xi_y - 2i\Omega_p \sigma \xi_x = -(c^2+v_A^2)ik_y
 \dfrac{\delta \rho}{\rho_0} - v_A^2 \left( -k_y k_z \xi_z + k_z^2 \xi_y
				     \right) - ik_y \psi_p, \\
 \label{pert_EoMz}
& -\sigma^2 \xi_z = -c^2 ik_z \dfrac{\delta \rho}{\rho_0} - ik_z \psi_p,
\end{eqnarray}
where $v_A^2=B_0^2/4\pi \rho_0$ denotes the \alfven\ velocity of the
background gas.  We shall also define, for later convenience, the plasma
$\beta$ by $c^2/v_A^2$.

The equations \eqref{pert_EoC}, \eqref{pert_EoMx}, \eqref{pert_EoMy} and 
\eqref{pert_EoMz} are four independent equations for four variables 
$\delta \rho$ and $\vect{\xi}$.  The boundary conditions to be imposed
are such that wave excited propagate away from the planet in both inner
and outer parts of the disk.

Once the wave pattern is derived for each Fourier mode, $z$-component of
the torque, of which backreaction causes the orbital migration, exerted
on the disk by the planet for each mode is calculated by
\begin{equation}
 T_{k_y,k_z} = - 2 L_y L_z \rho_0 r_p k_y 
  \int \mathrm{Im}\left(\dfrac{\delta
		      \rho_{k_y,k_z}(x)}{\rho_0}\right)
  \psi_{\rm{p}k_y k_z}(x) dx,
  \label{tq_calc}
\end{equation}
where $\mathrm{Im}$ denotes the imaginary part.

\subsection{Wave Propagation Property of the Disk}

We shall investigate the wave propagation property of the disk with
poloidal magnetic field.  First, we derive a wave
equation from \eqref{pert_EoC}-\eqref{pert_EoMz}.  The $x$- and
$y$-components of the equations of motion can be written
\begin{eqnarray}
\label{eqn1}
& (\sigma^2 + 3\Omega_p^2 - v_A^2 k_z^2)\xi_x -2i\Omega_p \sigma \xi_y =
 \dfrac{df}{dx}, \\
\label{eqn2}
& (\sigma^2 - v_A^2 k_z^2) \xi_y + 2i\Omega_p \sigma \xi_x = ik_y f,
\end{eqnarray}
where $f(x)$ is defined by
\begin{equation}
\label{fdef}
 f(x) \equiv \dfrac{1}{\sigma^2} \left[ \left\{ (c^2+v_A^2)\sigma^2 -
					 c^2 v_A^2 k_z^2 \right\}
 \dfrac{\delta \rho}{\rho_0} + (\sigma^2 - v_A^2 k_z^2) \psi_p \right].
\end{equation}
Equations \eqref{eqn1}
and \eqref{eqn2} are the generalization of equation (10) of
Goldreich \& Tremaine (1979).  Variable $f(x)$ is related to the
perturbation of total pressure
$\delta \Pi = c^2 \delta \rho + B_0 \delta B_z/4\pi$ by 
\begin{equation}
 f(x) = \frac{\delta \Pi}{\rho_0} + \psi_{\rm p}.
\end{equation}
Therefore, it is a natural extension of the variable used by Goldreich
\& Tremaine (1979) that is $c^2 \delta \rho/\rho_0 + \psi_{\rm p}$.
Solving for $\xi_x$ and $\xi_y$,
\begin{eqnarray}
& \xi_x = \dfrac{1}{D} \left[ \left( \sigma^2 - v_A^2 k_z^2  \right)
		      \dfrac{df}{dx} - 2\Omega_p \sigma k_y f \right], \\
& \xi_y = \dfrac{1}{D} \left[ -2i\Omega_p \sigma \dfrac{df}{dx} 
			+ \left( \sigma^2 + 3\Omega_p^2 - v_A^2 k_z^2
			  \right) ik_y f \right],
\end{eqnarray}
where $D$ is
\begin{equation}
 D = (\sigma^2 - v_A^2 k_z^2)(\sigma^2 - v_A^2 k_z^2 +3 \Omega_p^2) -
  4\sigma^2 \Omega_p^2.
\end{equation}
This generalizes what is denoted by $D$ in the case of unmagnetized
disk, e.g., equation (12) of Goldreich \& Tremaine (1979).  Note
that in the absence of magnetic field, $D$ is a quadratic function of
$\sigma$, while this becomes a quartic function in the present
situation.  From equations \eqref{pert_EoC} and \eqref{pert_EoMz},
we finally
obtain a second order ordinary differential equation which describes
wave excitation and propagation of the disk,
\begin{equation}
 \dfrac{d^2 f}{dx^2} + \mathcal{A}_1 \dfrac{df}{dx} +\mathcal{A}_2 f =
  \mathcal{S},
  \label{eqn_prop}
\end{equation}
where
\begin{eqnarray}
\label{a1}
& \mathcal{A}_1 = \dfrac{d}{dx} \ln \dfrac{\sigma^2-v_A^2 k_z^2}{D}, \\
\label{a2}
& \mathcal{A}_2 = \dfrac{(\sigma^2 - c^2 k_z^2)D}{\left\{
		(c^2+v_A^2)\sigma^2 - c^2 v_A^2 k_z^2\right\}
                (\sigma^2 - v_A^2 k_z^2)} 
	+ \dfrac{2\Omega_p \sigma k_y}{\sigma^2 - v_A^2 k_z^2} 
	\dfrac{d}{dx}(\ln D) - k_y^2, \\
\label{sourceS}
& \mathcal{S} = \dfrac{\sigma^2 D}{\left\{ (c^2+v_A^2)\sigma^2 - c^2 v_A^2
			k_z^2 \right\} (\sigma^2 - v_A^2 k_z^2)} \psi_p .
\end{eqnarray}

\begin{figure}
 \plotone{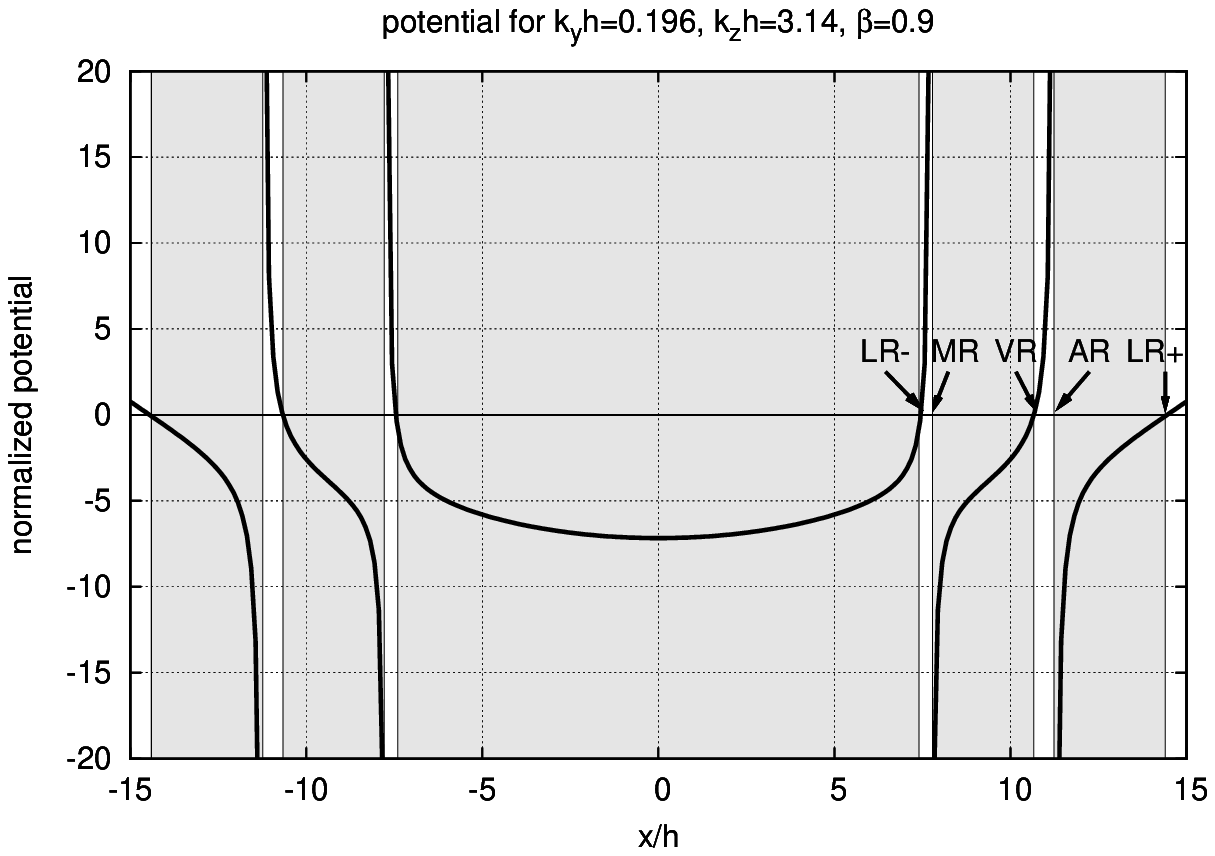}
 \figcaption{Potential $V(x)$ given by equation \eqref{potV} for
 $\beta=0.9$, $k_yh=0.196$, and $k_zh=3.14$.  Resonance positions in the
 outer disk ($x>0$) are 
 indicated.  LR$+$ and LR$-$ denote Lindblad resonances, AR denotes
 \alfven\ resonance, VR denotes vertical resonance, and MR denotes
 magnetic resonance.  The grey regions correspond to the evanescent
 regions.  Note that regions $|x/h|>15$ are all propagation regions.
 \label{fig:pot} }
\end{figure}

Imposing WKB approximation, $\ df/dx, \  k_z f \gg k_y f $,
this equation simplifies to \\ Schr\"{o}dinger type:
\begin{equation}
 \dfrac{d^2 f}{dx^2} + V(x) f = \mathcal{S}
\end{equation}
where
\begin{equation}
 V(x) = \dfrac{(\sigma^2 - c^2 k_z^2)D}{\left\{
		(c^2+v_A^2)\sigma^2 - c^2 v_A^2 k_z^2\right\}
                (\sigma^2 - v_A^2 k_z^2)}.
\label{potV}
\end{equation}
The regions where $V(x)>0$ are wave propagation regions, and those where
$V(x)<0$ are evanescent.  The boundary between these regions, where
$V(x) = 0$ or $V(x) = \pm \infty$, is the resonances.  Figure
\ref{fig:pot} shows the appropriately normalized potential $V(x)$ for
disk with $\beta=0.9$ and mode $k_yh=0.196$ and $k_zh=3.14$.

There are two or three points where $V(x) = 0$ in one side of the disk
with respect to the planet (either $x>0$ or $x<0$), depending on the
value of $\beta$.  From the condition $D=0$, we have
\begin{equation}
 \sigma^2 = v_A^2 k_z^2 + \dfrac{1}{2} \left[ \Omega_p^2 \pm
				      \sqrt{\Omega_p^4 + 16 \Omega_p^2
				      v_A^2 k_z^2} \right]
\end{equation}
We call the point with positive sign Lindblad Resonance Plus (LR$+$)
and with negative sign Lindblad Resonance Minus (LR$-$).  In the unmagnetized
disk, LR$+$ coincides with the usual Lindblad resonance, 
$\sigma^2 = \kappa^2$, where $\kappa$ is the epicycle frequency, and LR$-$
degenerates into corotation point.  We note that LR$-$ exists
only when $v_A^2 k_z^2 > 3\Omega_p^2$, which is exactly the same as the
stability condition against  MRI \cite{BH91}.  When LR$-$ does not exist but
$k_z \neq 0$, the corotation 
region becomes a wave propagation region.  Since $\sigma=0$ at the
corotation, this indicates that there is a mode with zero frequency but
non-zero wavelength, and therefore, in general, there is an unstable
mode.  Another condition for $V(x)=0$ is
\begin{equation}
 \sigma^2 = c^2 k_z^2 .
\end{equation}
This condition does not depend on magnetic field strength.  This resonance
corresponds to that found by Takeuchi \& Miyama (1998) and is named
``Vertical Resonance'' by Zhang \& Lai (2006).  We shall also call this
point Vertical Resonance (VR) in this paper.

There are two points in one side of the disk where $V(x)$ diverges.
One is given by
\begin{equation}
 \sigma^2 = v_A^2 k_z^2 .
\end{equation}
At this point, the radial wavelength of \alfven\ wave becomes zero.  We 
shall call this point \alfven\ Resonance (AR).  Note that this
divergence is related to what is called \alfven\ resonance in plasma
physics (see e.g., Stix 1992).
The other point where $V(x)$ diverges is given by
\begin{equation}
 \sigma^2 = \dfrac{c^2 v_A^2 k_z^2}{c^2 + v_A^2}.
\end{equation}
At this point, the wavelength of the slow mode becomes zero, and this
corresponds to ``Magnetic Resonance'' found by Terquem (2003) in the
analysis of toroidal field.  Therefore, we shall also call this point
Magnetic Resonance (MR) in this paper.

\begin{figure}
 \plotone{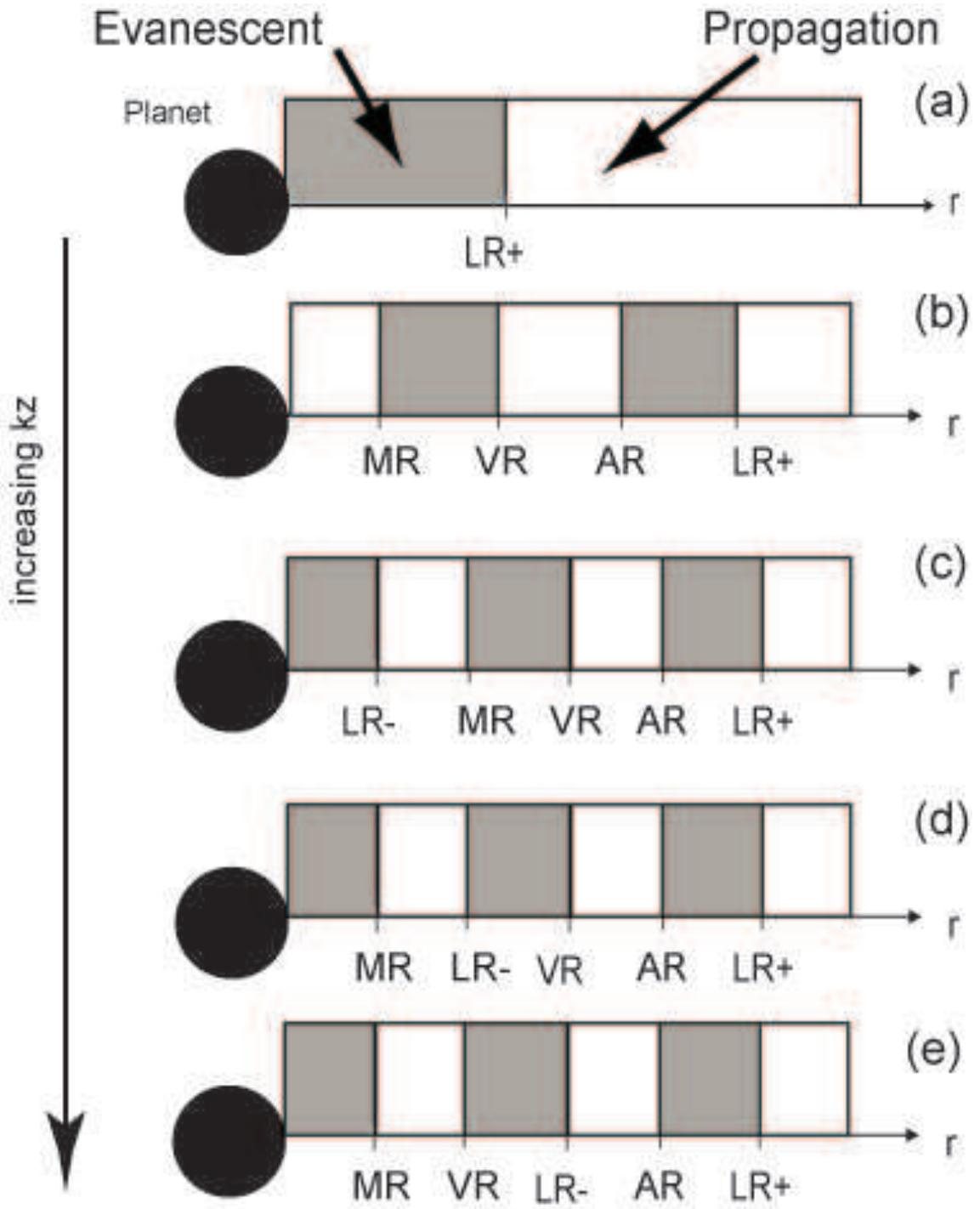}
 \figcaption{Wave propagation property for $\beta<1$ disk.  The
 parameters used in figure \ref{fig:pot} corresponds to case (c).
  \label{fig:waveprop}}
\end{figure}

For two-dimensional modes, $k_z=0$, only LR+ exists, and the region in
the vicinity of the corotation is evanescent region, whereas the regions
further away from LR+ are propagation regions.  The details of the wave
propagation property of the disk depend on the value of $\beta$ and
$k_z$, but in general, there are three propagation regions on one side
of the corotation, corresponding to the three wave modes of 
the magnetohydrodynamics.  
There are only two propagation regions in the analysis of Terquem
(2003), since the mode is restricted to $k_z=0$ and the \alfven\ wave
with $\delta v_z \neq 0$ is not taken into account.  
Figure \ref{fig:waveprop} shows the wave propagation property of one
side of the disk in the case of $\beta<1$.

With careful investigation of the properties of the resonances, it is
possible to analyze the wave excitation, propagation, and
absorption.  In this paper, however, we calculate the torque in some
restricted cases and compare the 
results with numerical calculation.  We first investigate the case when
$k_z=0$.  We then calculate three dimensional modes, $k_z \neq 0$, when
very strong magnetic field is exerted.  In these cases, MRI does not
occur \cite{BH91}, and we expect the wave pattern becomes stationary
with respect to the planet's motion.

\subsection{Two Dimensional Mode, $k_z=0$}

We consider the two-dimensional, or $k_z=0$, mode.  In this case it is
possible to calculate the torque exerted on the disk without imposing
WKB approximation.  When $k_z=0$, it is clear from equation
\eqref{pert_EoMz} that the fluid particles do not
move along $z$-axis, $\xi_z=0$, and the effect of magnetic field appears
only in the pressure term.  The sound speed becomes the phase velocity
of the fast mode, $c^2 + v_A^2$, and this acts as an effective sound
speed.  We can then follow the track of Artymowicz (1993) and calculate
the torque exerted on the disk by evaluating the angular momentum flux
carried by the wave at $|x|\to \infty$. 

The position of the effective Lindblad resonance is given by
\begin{equation}
 \sigma^2(x_{\rm eff}) - \Omega_p^2 - c^2 k_y^2 (1 + \beta^{-1}) = 0 .
\end{equation}
When
\begin{equation}
 \dfrac{\Omega_p^2 + c^2 k_y^2 (1+\beta^{-1})}{\Omega_p ck_y
  \sqrt{1+\beta^{-1}}} \gg 1 ,
  \label{cond:ART}
\end{equation}
the resonances inside and outside of the corotation radius are well isolated
each other, and the torque exerted on the disk may be evaluated by the
strength of the gravitational potential of the planet at the resonance
point.  The modified formula of the torque is
\begin{equation}
 T_{\rm 2D} = \dfrac{2\pi}{3} r_p \rho_0 L_y L_z \dfrac{\Omega_p}{\Omega_p^2 + 4
  c^2 k_y^2 (1+\beta^{-1})} \dfrac{1}{\sqrt{\Omega_p^2 + c^2
  k_y^2(1+\beta^{-1})}} \Psi^2_{\rm eff},
  \label{tq_ART}
\end{equation}
where
\begin{equation}
 \Psi_{\rm eff} = \frac{d \psi_p}{dx}(x_{\rm eff}) -
  2 k_y \dfrac{\sqrt{\Omega_p^2 + c^2 k_y^2 (1+ \beta^{-1})}}{\Omega_p}
  \psi_p(x_{\rm eff}) .
\end{equation}
Details of the derivation are given in Appendix \ref{app:2dtq}.

The equation \eqref{tq_ART} generalizes the equation (56) of Artymowicz
(1993) to a magnetized disk.  Since the effective sound speed becomes
faster when poloidal magnetic fields present, the magnitude of the
torque becomes smaller.  There are two
kinds of cutoff mechanism of the torque, as Artymowicz (1993) pointed
out. 
One is mild cutoff that comes from the coefficient of $\Psi_{\rm eff}$.
The other 
is the sharp cutoff, which is the consequence of the fact that position
of the effective Lindblad resonance goes further away from the
corotation when magnetic field is stronger.  In the case of planetary
migration, the effect of sharp cutoff is more important, and the torque
by two-dimensional mode is 
strongly suppressed when $\beta \lesssim 1$.  We emphasize that we have
obtained the torque formula \eqref{tq_ART} including the torque cutoff
at high $k_y$ mode because we have not imposed WKB
approximation, and hence, not neglected terms with $k_y$.

\subsection{Three Dimensional Mode, $k_z \neq 0$, in the Limit of Strong
  Magnetic Field}

In the unmagnetized disk, the contribution from three-dimensional, or
$k_z \neq 0$, mode is small \cite{TTW02}.  However, it is indicated in
the previous section that the two-dimensional mode is strongly suppressed
when strong magnetic field is present.  Therefore three dimensional mode
may be important in this case.  We investigate three-dimensional mode in
the limit of $\beta=0$.  We impose WKB approximation in this section.

In the limit of $\beta\to 0$, the resonances LR+, LR-, and AR are
infinitely far away from the corotation and we can safely neglect their
contribution.  The other two resonances, MR and VR, degenerate
and the equation \eqref{eqn_prop} becomes
\begin{equation}
 \dfrac{d^2f}{dx^2} - k_z^2 f = S(x),
\end{equation}
where the source term $S(x)$ is
\begin{equation}
 S(x) = - \dfrac{\sigma^2}{\sigma^2 - c^2 k_z^2} \psi_p .
\end{equation}
In this case, waves are evanescent in the vicinity of the corotation,
but there is a singularity in the source term at the
degenerate point of MR and VR,
\begin{equation}
 \sigma^2_{\rm MR} = c^2 k_z^2 .
\end{equation}
The subscript MR denotes the quantities
evaluated at this point.
Note that this divergence originally comes from the divergence of the
source term \eqref{sourceS} at MR.

In order to regularize the singularity, we consider the small
viscosity effective only in the vicinity of this point.  The viscosity
is effectively taken into account by adding the small positive
imaginary part to the frequency,
\begin{equation}
 \sigma(x) \to \sigma(x) + i\epsilon,
\end{equation}
where $\epsilon >0$ is small positive number (see e.g., Meyer-Vernet
and Sicardy 1987).  Taking the limit of
$\epsilon \to 0$,
\begin{equation}
 \int \dfrac{g(x)}{\sigma - ck_z} dx = \mathcal{P}\int
  \dfrac{g(x)}{\sigma-ck_z} dx -i\pi \int 
  \delta_D(\sigma-ck_z) g(x) dx
\end{equation}
where $g(x)$ is an arbitrary smooth function, $\mathcal{P}$ denotes the
principal value of the integration and 
$\delta_D(x)$ is the Dirac's delta function.

The boundary condition we impose is that the perturbation must vanish for
$|x|\to \infty$, i.e.
\begin{eqnarray}
& f(x \to \infty) \propto e^{-k_z x}, \\
& f(x \to -\infty) \propto e^{k_z x}.
\end{eqnarray}
The solution that satisfies this condition is
\begin{equation}
 f(x) = - \dfrac{1}{2k_z} \left[ e^{-k_z x} \int_{-\infty}^{x} e^{k_z
			   u} S(u) du - e^{k_z x} \int_{+\infty}^{x}
			   e^{-k_z u} S(u) du  \right] .
\end{equation}
Substituting $S(x)$, the real and imaginary part of the solution are
\begin{eqnarray}
 \dfrac{2}{k_z} \mathrm{Re} f = \mathcal{P} \int_{-\infty}^{x}
  e^{k_z(u-x)} \psi_p(u) 
  \left[ 1 + \dfrac{ck_z}{2} \left( \dfrac{1}{\sigma(u)-ck_z}
			      - \dfrac{1}{\sigma(u)+ck_z} \right)
  \right] du  \nonumber \\
 -  \mathcal{P} \int_{+\infty}^{x} e^{k_z(x-u)} \psi_p(u)
  \left[ 1 + \dfrac{ck_z}{2} \left( \dfrac{1}{\sigma(u)-ck_z}
			      - \dfrac{1}{\sigma(u)+ck_z} \right)
  \right] du
  \label{realf}
\end{eqnarray}
and
\begin{eqnarray}
 \label{imf}
 - \dfrac{6\Omega_p k_y}{\pi c k_z^2 \psi_{p,\rm MR}} \mathrm{Im} f =
  \left\{
  \begin{array}{cc}
   e^{k_zx} \left[ \exp\left( -\dfrac{2ck_z^2}{3\Omega_pk_y} \right) -
	      \exp\left( \dfrac{2ck_z^2}{3\Omega_pk_y} \right)
	      \right] & x<-x_{\rm MR}  \\[15pt]
   \exp\left[ -\dfrac{2ck_z^2}{3\Omega_pk_y} \right] \left( e^{-k_z x} -
	e^{k_z x}\right) & -x_{\rm MR} < x < x_{\rm MR} \\[15pt]
   - e^{-k_zx} \left[ \exp\left( -\dfrac{2ck_z^2}{3\Omega_pk_y} \right) -
	      \exp\left( \dfrac{2ck_z^2}{3\Omega_pk_y} \right)
	      \right] & x>x_{\rm MR}
  \end{array}
  \right.
\end{eqnarray}

In the limit of $\beta \to 0$, the density perturbation is given by 
\begin{equation}
 \dfrac{\delta \rho}{\rho_0} = \dfrac{1}{\sigma^2 - c^2 k_z^2} 
  \left[ \dfrac{\sigma^2}{v_A^2}f + k_z^2 \psi_p \right] .
 \label{rho_f}
\end{equation}
In order to calculate the torque, we need imaginary part of
 $\delta \rho / \rho_0$ which is given by
\begin{eqnarray}
 2ck_z \mathrm{Im}\left[ \dfrac{\delta \rho}{\rho_0} \right] = - \pi
  \left\{ \delta_D(\sigma - ck_z) - \delta_D(\sigma + ck_z) \right\} 
 \left( \dfrac{\sigma^2}{v_A^2} \mathrm{Re}f + k_z^2 \psi_p  \right)
 \nonumber \\
 + \dfrac{\sigma^2}{v_A^2} \left\{ \dfrac{\mathcal{P}}{\sigma - ck_z} 
		- \dfrac{\mathcal{P}}{\sigma + ck_z} \right\}
 \mathrm{Im}f .
\end{eqnarray}
It is possible to show that $f$ does not diverge at the resonance
(see Appendix \ref{app:realf} for details), and we can neglect the term
with $\sigma^2/v_A^2$ in right hand side of the equation \eqref{rho_f}
when magnetic field is strong enough.  Quantitatively,
we can neglect these terms when
\begin{equation}
 \dfrac{\sigma^2}{v_A^2} \dfrac{ck_z^2}{\Omega_p k_y} \ll k_z^2 
\end{equation}
since $f\sim\mathcal{O}(ck_z^2\psi_p/\Omega_p k_y)$.  Since 
$\sigma^2\sim c^2 k_z^2$ in the vicinity of the resonance, we obtain
\begin{equation}
 \beta \ll \dfrac{\Omega_p k_y}{c k_z^2}.
  \label{fieldcond}
\end{equation}
When this condition is satisfied, the imaginary part of the density
perturbation is
\begin{equation}
 \mathrm{Im} \left[ \dfrac{\delta \rho}{\rho_0} \right] \sim -\dfrac{\pi
  k_z}{2c} 
  \left\{ \delta_D(\sigma-ck_z) - \delta_D(\sigma+ck_z) \right\} \psi_p
  .  
\end{equation}
The first delta function indicates the torque exerted on the outer disk,
and the second inner disk.  These torques are of the same magnitude but
different in sign.  The magnitude of the torque on one side of the disk
is then, from \eqref{tq_calc},
\begin{equation}
 T_{\rm MR} = \dfrac{2\pi}{3} L_y L_z \dfrac{ \rho_0 r_p k_z}{\Omega_p
  c} \psi_{p,\rm MR}^2 . 
  \label{tq_MR}
\end{equation}

\section{Numerical Calculation} \label{simu}

We have performed numerical calculations in order to investigate how well
the equations \eqref{tq_ART} and \eqref{tq_MR} describe the realistic
value of the torque.  We have done two sets of runs. One is for a 
two-dimensional disk.  The other is for a three-dimensional thick disk.

\subsection{Numerical Methods}

We adopt the nested grid method (see, e.g., Machida et al. 2005,
  Matsumoto \& Hanawa, 2003)
  to obtain high spatial resolution near the planet.
  Each level of rectangular grid has the same number of cells 
 ($ = 64 \times 256 $) for 2D run,
  while ($ = 64 \times 256 \times 16 $) for 3D run.  
  The cell width $\Delta s(l)$ depends on the grid level $l$.
 The cell width is reduced by half with increasing
  grid level ($l \rightarrow l+1$).
  We use 4 grid levels ($l=$1,2 $\cdot \cdot \cdot$ 4) for 2D run and 5
  levels for 3D run.
 The box size of the  coarsest grid $l=1$ is chosen to 
 $(L_x,L_y) = (64h, 256h)$ for 2D run and
$(L_x,L_y,L_z/2) = (64h, 256h, 16h)$ for 3D run.
  Note that in $z$-direction,
 the simulation box extends from midplane to $z=L_z/2$.
 The box size of the finest grid is $(x,y) = (2h, 8h)$ for 2D
  run and $(x,y,z) = (2h, 8h,h )$ for 3D run.
 The cell width of the coarsest grid is $\Delta s(1) = h$, while
  that of the finest grid has  $\Delta s(4)= 0.125h$ for 2D run and
  $\Delta s(5) = 0.0625h$ for 3D run. 
 We assume the fixed boundary condition in the $x$-direction and
  periodic boundary condition in the $y$-direction.  For $z$-direction,
  we impose a periodic boundary condition between $z=-L_z/2$ and
  $z=L_z/2$.

For two dimensional calculation, we neglect the $z$-dependence of the
planet potential, i.e. we adopt the potential of the form
\begin{equation}
 \psi_p = - \dfrac{GM_p}{\sqrt{x^2 + y^2}}.
\end{equation}

We employ the softening in the gravitational potential as follows.  The
gravitational force $\vect{F}$ by the planet is given by
\begin{equation}
 \vect{F} = \frac{GM_p}{(r + \varepsilon)^3} \vect{x},
\end{equation}
where $\varepsilon$ is the softening length, $r$ is the distance from
the planet's position, and $\vect{x}$ is the position vector.  We choose 
$\varepsilon$ such that this equals the mesh size of the finest grid,
i.e., $\varepsilon = 0.125h$ for 2D run and $\varepsilon=0.0625h$ for 3D
run.

We fix the planet mass to be $\tilde{r}_{\rm H} = 0.3$, corresponding to
$3 M_{\oplus}$ planets when 
$M_c=M_{\odot}$ and $h/r_p=0.05$, and vary the
initial strength of the poloidal magnetic field.
We performed the
calculations for $\beta=\infty$, 100, 10, 2, 0.3, 0.1, 0.01, and 0.001.

\section{Comparison between Numerical Calculation and Linear Analysis}
\label{compare}

For all two-dimensional calculations and for
three-dimensional calculation with $\beta=\infty$, 0.01, and 0.001,
we do not observe MRI and steady states are realized.  This is
consistent with the stability criterion of MRI derived from linear
analysis.

We then Fourier transform the density pattern of the steady state in
$y$- and $z$-directions and calculate the torque exerted on one side of
the disk by equation \eqref{tq_calc}, and this torque is compared
with the results of linear analysis.  The normalization of the torque is
taken to be
\begin{equation}
 \tilde{T} = \dfrac{T}{\rho_0 r_p H^4 \Omega_p^2} .
\end{equation}

We make use of FFT (e.g., Press et al. 1992).  The
wavenumber $k$ we evaluate is given by
\begin{equation}
 k = \dfrac{2\pi n}{L} ,
\end{equation}
where $L$ is the box size of the $y$- or $z$-directions and $n$ is an
integer with $-N/2<n<N/2$ where $N$ is the mesh number.  We also Fourier
transform the gravitational potential of the planet numerically to
obtain the value of $\psi_p$.

\subsection{Two-dimensional Calculation}

\begin{figure}
 \plotone{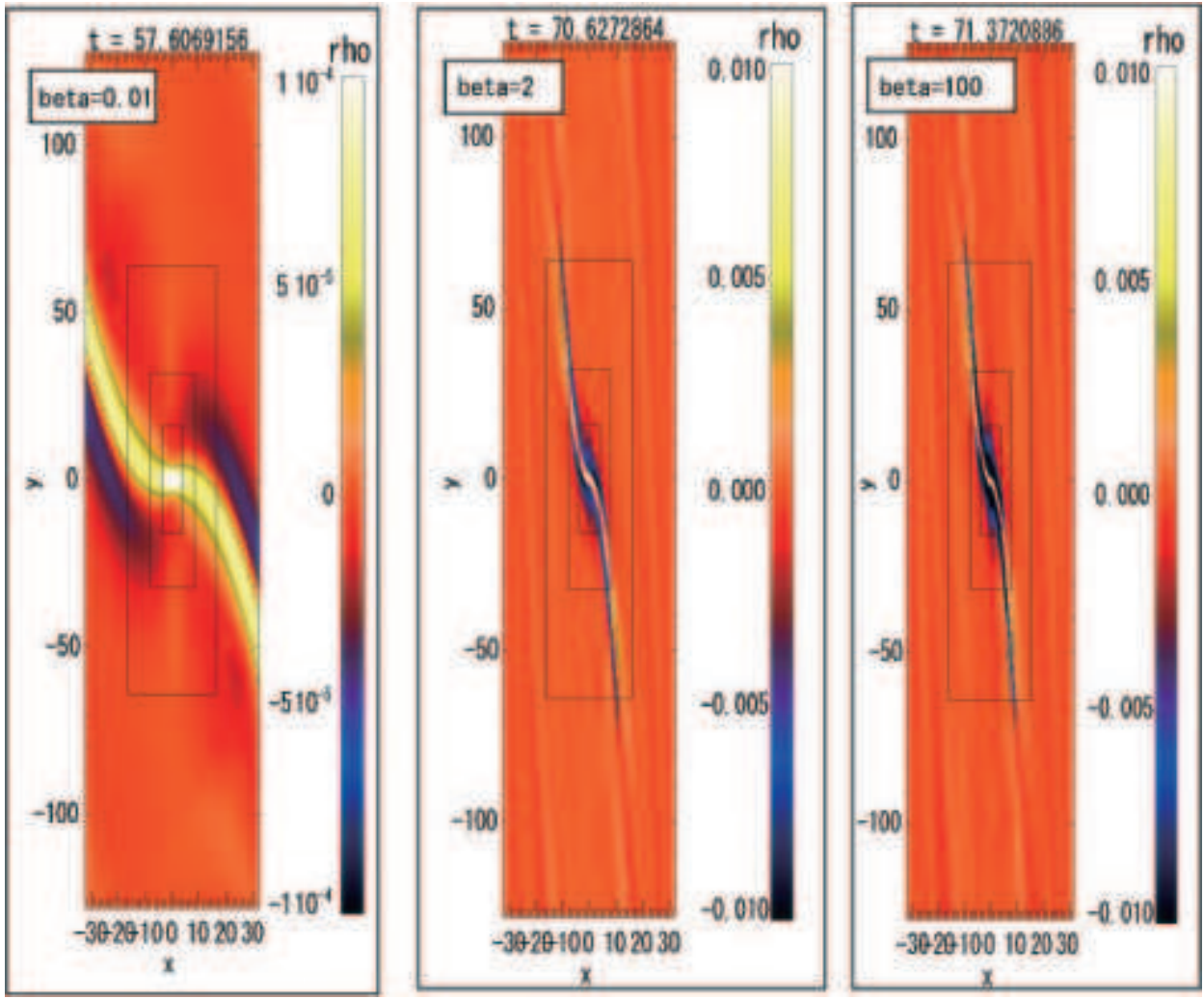}
 \epsscale{.80}
 \figcaption{Density profile obtained by the two-dimensional
 numerical calculation.  The pattern of density perturbation
 $\delta \rho / \rho_0$ is indicated by false color.  It is clear that
 the stronger the magnetic 
 field, the further the point where waves are excited and the smaller the
 amplitude.  Note that color scales are different for different values
 of $\beta$. The $x$- and $y$-axes correspond to the axes of
 shearing-sheet, normalized by the scale height $c/\Omega_p$. The
 elapsed time $t$ is normalized by the planet's Kepler time
 $\Omega_p^{-1}$.  Four different levels of nested grid are super-imposed.
 \label{fig:2dsimu} }
\end{figure}

\begin{figure}
 \plotone{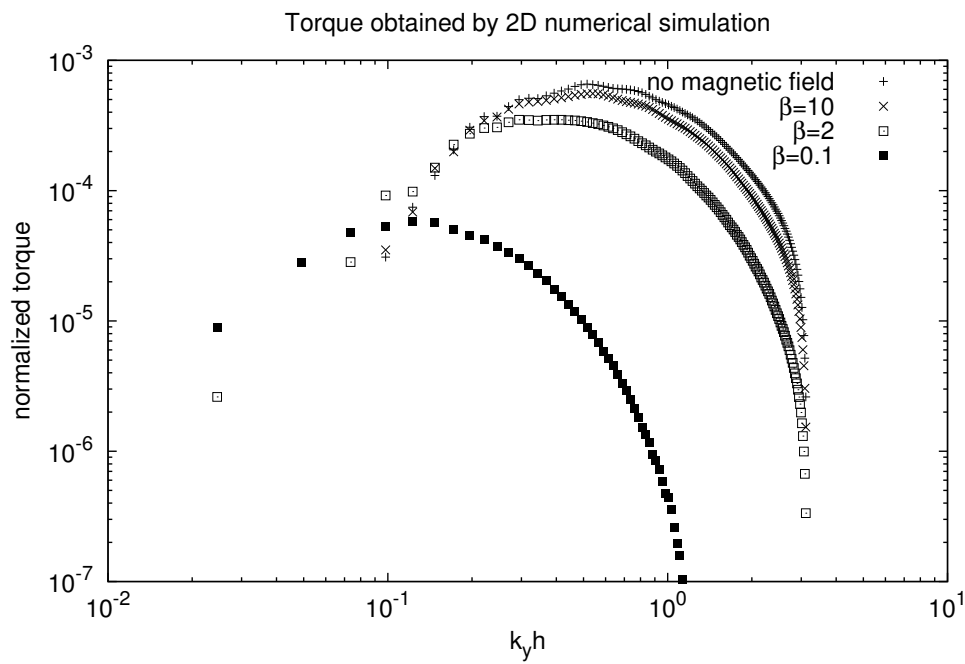}
 \figcaption{The dependence of two-dimensional torque ($k_z=0$ modes) on
 the strength of magnetic field obtained by numerical calculations.  The
 models with $\beta=\infty$ (no magnetic field, plus), 10 (cross), 2
 (open square), and 0.1 (filled square) are shown.  The
 torque is cut off drastically for models with $\beta<1$.  The
 horizontal axis denotes $k_yh$ and the vertical axis denotes normalized
 torque. 
 \label{fig:2dtq_simu} }
\end{figure}

Figure \ref{fig:2dsimu} shows the stationary pattern of density
perturbation obtained by
two dimensional calculations for $\beta=0.01$, $2$, and $100$. 
It is clear 
that, with increasing magnetic field, the amplitude of the wave becomes
small and the point where waves are excited goes further away from the
planet.  Figure \ref{fig:2dtq_simu} shows the torque calculated as a
result of numerical calculation for various magnetic field strength, or
different $\beta$.  It
is clear that the torque becomes weaker as the magnetic field is
stronger.

\begin{figure}
 \plotone{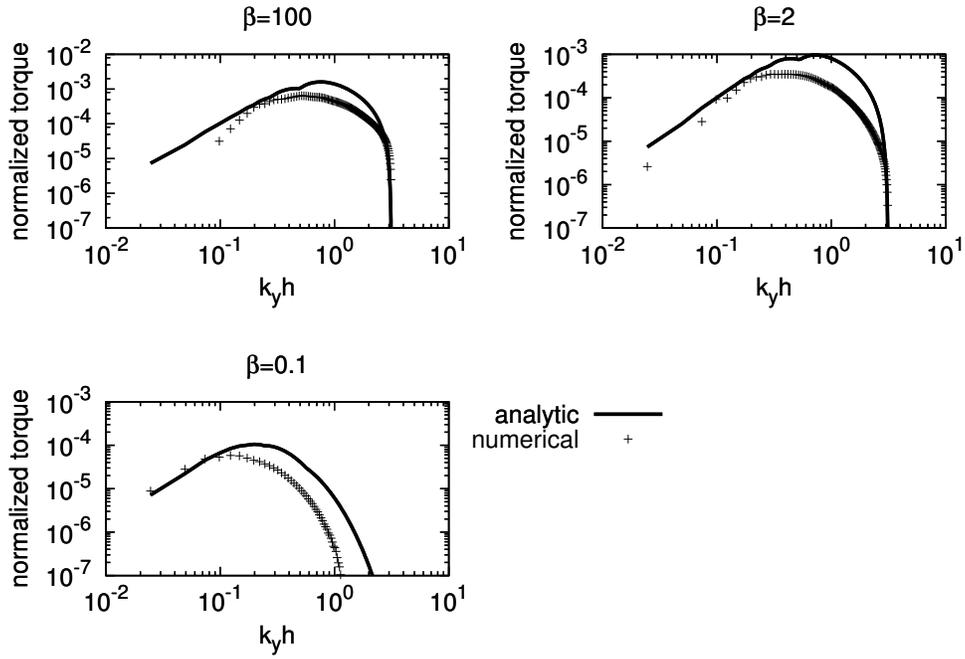}
 \figcaption{Comparison of the torque obtained by the two-dimensional
 numerical calculation (plus) and the linear analysis (line), the
 equation \eqref{tq_ART} for $\beta=100$ (top left), $\beta=2$ (top
 right), and $\beta=0.1$ (bottom).  
 The horizontal axis denotes the azimuthal mode number
 and the vertical axis denotes normalized torque.
 \label{fig:2dtq_compare} }
\end{figure}

We show in figure \ref{fig:2dtq_compare} the comparison between the
results of numerical 
calculation and linear analysis, equation \eqref{tq_ART}.  It is
clear that for modes that satisfy condition \eqref{cond:ART}, which
we expect that equation \eqref{tq_ART} gives a good approximation
for the torque, numerical calculation and linear analysis indeed show
reasonably good agreement, at least an order of magnitude, even though
equation \eqref{tq_ART} estimates the torque by the value of
density perturbation only at the position of effective Lindblad
resonance.  Therefore, equation \eqref{tq_ART} is useful for 
estimating two-dimensional torque when poloidal magnetic field is
exerted on the disk.

We also checked that the numerical calculation and linear analysis are
in good agreement for other values of $\beta$ except for $\beta=0.001$.
For $\beta=0.001$, since the amplitude of density perturbation is very
small, numerical torque is dominated by small noise in the disk.

\subsection{Three-dimensional Calculation}

\begin{figure}
 \plotone{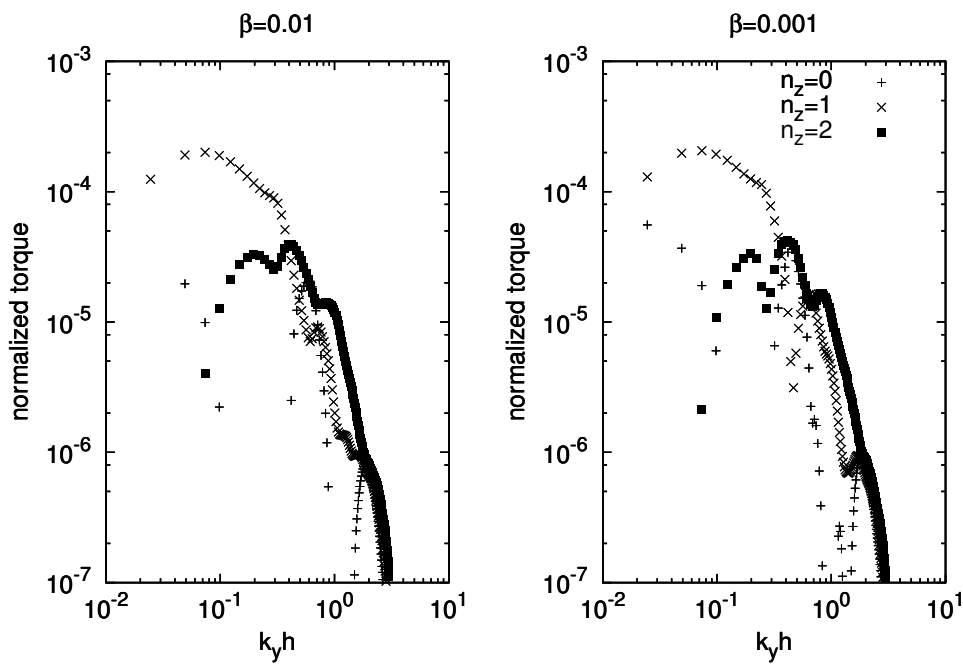}
 \figcaption{$k_z$-dependence of the torque obtained by three-dimensional
 numerical calculations.  The left panel shows $\beta=0.01$ and the
 right $\beta=0.001$.  The horizontal axis denotes the azimuthal mode
 number and the vertical axis denotes normalized torque.
 Two-dimensional modes are denoted by plus, three-dimensional modes with
 $n_z=1$ by cross, and 
 $n_z=2$ by filled square.  Three-dimensional modes with $n_z=1$
 dominate the two-dimensional modes when the magnetic field is
 sufficiently strong ($\beta \ll 1$).
 \label{fig:3dtq_num} }
\end{figure}

For $\beta=0.01$ and $\beta=0.001$ models of the three-dimensional
calculations, we do not observe 
MRI and steady state is realized.
For other parameters, we observe the 
instability.  Since we investigate the stationary pattern, we focus on
results in which we do not observe MRI.
We show in figure \ref{fig:3dtq_num}
the torque that is derived from numerical calculation for
 $n_z=0,1, \rm{and}\ 2$,
where $n_z$ is the mode number of $z$-direction.
It is clear that $n_z=1$ modes overwhelm the two-dimensional modes in
these models.

\begin{figure}
 \plotone{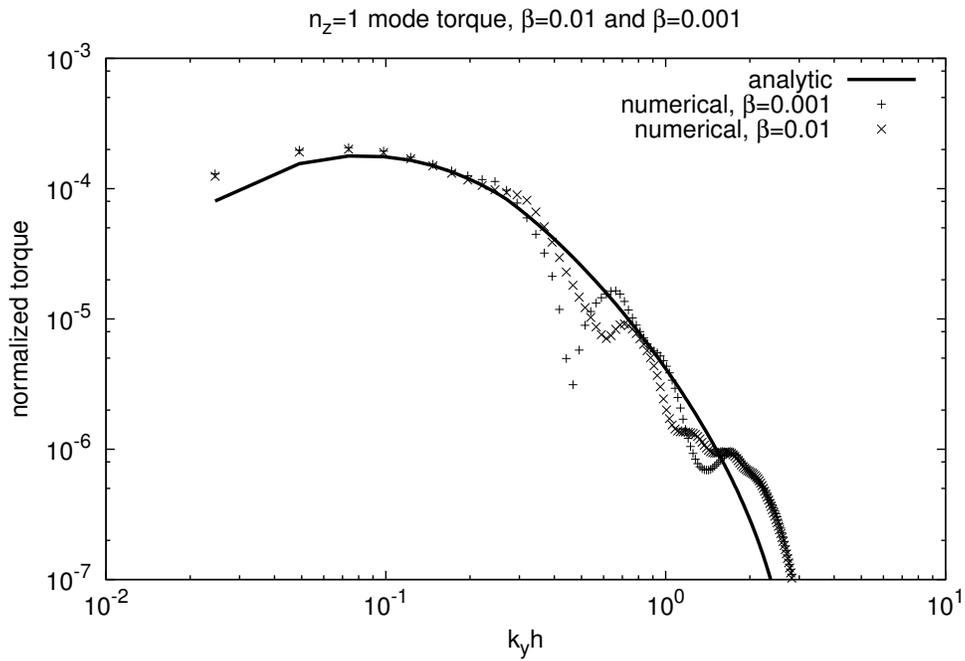}
 \figcaption{The comparison of $n_z=1$ mode torque between the analytic
 formula (line), equation \eqref{tq_ART}, and 
 three-dimensional numerical calculation for $\beta=0.001$ (plus) and
 $\beta=0.01$ (cross).  The horizontal axis  shows the azimuthal mode
 number and the vertical axis shows the normalized torque.
 \label{fig:3dtq_comp} }
\end{figure}

Figure \ref{fig:3dtq_comp} compares the torque calculated from the
three-dimensional numerical
calculations and that calculated from linear analysis of $n_z=1$ modes,
torque formula \eqref{tq_MR}.  
From the derivation of formula \eqref{tq_MR}, this expression of the
torque is valid
when WKB condition $k_y \ll k_z$ and strong magnetic field condition
\eqref{fieldcond} are both satisfied.  In the present parameter, the WKB
condition is more restrictive.  Since $k_z h=2\pi/32=0.196$ for $n_z=1$
mode, we expect
that for $k_y h$ greater than this value, equation \eqref{tq_MR}
does not give a good approximation for the torque.  Nevertheless, the
result of the numerical calculation indicates that the equation
\eqref{tq_MR} shows a very good agreement even in the modes with $k_y h$
greater than this limit.

\begin{figure}
 \plotone{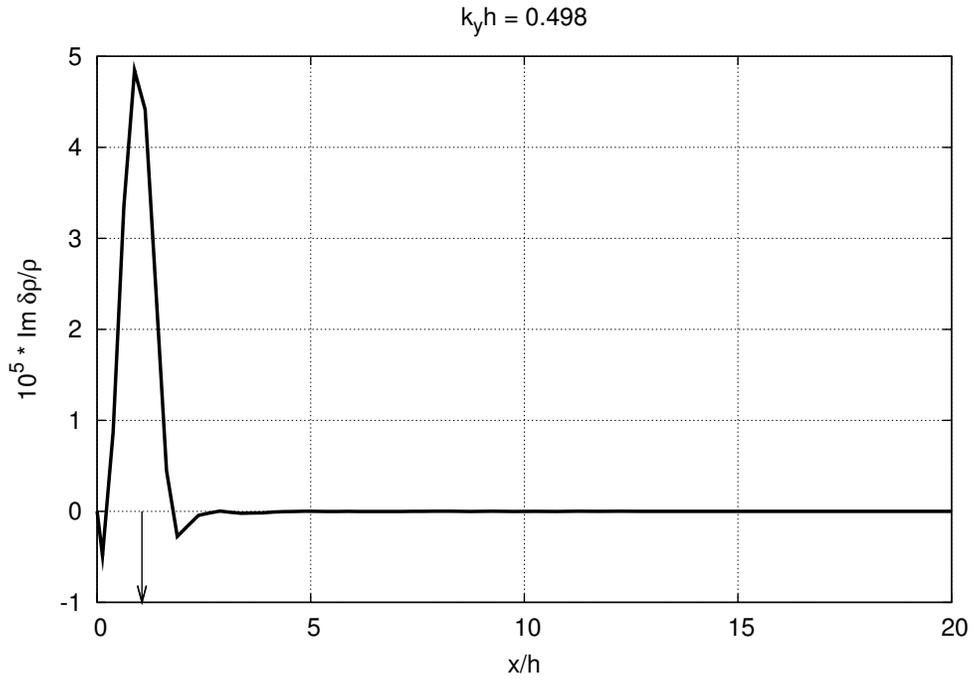}
 \figcaption{The radial profile of $\rm{Im} (\delta \rho/\rho_0)$ for
 the Fourier component $(k_yh, k_zh)=(0.498,0.785)$ obtained by
 $\beta=0.001$ numerical calculation.  The profile of the torque on the
 disk depends on the imaginary part of the density perturbation [see
 equation \eqref{tq_calc}].  
  The horizontal axis shows the radial
 coordinate and the arrow indicate the position of the magnetic
 resonances calculated by the linear analysis.
 \label{fig:denspert_3d} }
\end{figure}

We also find that the imaginary part of the Fourier components of
density perturbation 
diverges around MR, as expected from linear analysis.
Figure \ref{fig:denspert_3d} shows the profile of the imaginary part of
the density perturbation of $\beta=0.001$ calculation for 
$(k_yh, k_zh)= (0.498, 0.785)$.
  The position of magnetic resonance is indicated by
an arrow.  It is clear that density perturbation diverges at the
resonance position and the contribution of the torque
mostly comes from this divergence.  The torque is localized at the magnetic
resonances since waves cannot propagate on the disk, and,
therefore, the analytic torque formula 
\eqref{tq_MR} gives a good approximation of the total torque, even if
we consider the regions only in the vicinity of the resonance.

\section{Discussion} \label{discussion}

\subsection{The Strength of Three-dimensional Modes in a Thin Disk}

Tanaka et al. (2002) has shown that in the unmagnetized disk, three-dimensional
modes are subdominant.  In contrast, when poloidal
magnetic field is exerted on the disk, it is indicated that
three-dimensional modes can dominate the torque when the magnetic field
is sufficiently strong.  In this section, we briefly discuss the
critical value of $\beta$ at which $k_z \neq 0$ modes dominate the total
torque in a thin magnetized disk according to the results of
linear analysis.  By ``thin disk'', we refer to the disk with small aspect
ratio, smaller than that we have used in the numerical calculation, but
not two-dimensional.

\begin{figure}
 \plotone{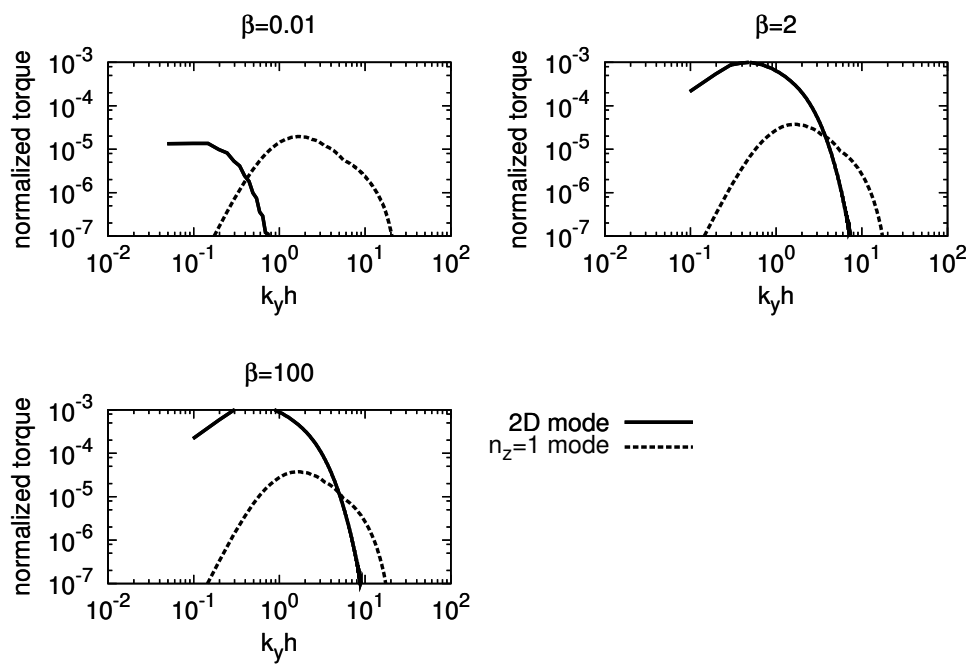}
 \figcaption{Comparison between two-dimensional torque formula
 \eqref{tq_ART} and three-dimensional formula \eqref{tq_MR} for thin
 disk with $\beta=0.01$ (top left), $\beta=2$ (top right), and
 $\beta=100$ (bottom). Two-dimensional torque is denoted by the solid
 line and three-dimensional torque is denoted by the dashed line.  The
 horizontal axis denotes azimuthal mode number
 and the vertical axis denotes normalized torque.
 \label{fig:thin_2d} }
\end{figure}

We calculate the torque for $n_z=0$ modes by \eqref{tq_ART} and
 $n_z = 1$ modes by \eqref{tq_MR}.  The Fourier transformation of the
 planet's gravitational potential is done numerically with
 the box size $L_x=32h, L_y=128h, L_z/2=2h$ and the spatial resolution
 $0.125h$.  We have checked that $n_z=2$ modes are smaller than
 $n_z=1$ modes.  Figure \ref{fig:thin_2d} shows the torque for 
 $\beta=0.01$, $2$, and $100$.  The vertical wavenumber of
 $n_z=1$ modes is $k_zh=1.57$, and WKB approximation is valid 
 for $k_yh$ less than this value.  For thin disk case, it is
 indicated, just as thick disk case, that two-dimensional modes are
 dominant for a weak magnetic field, while three-dimensional modes are
 important for a strong magnetic field case.  We have investigated 
 other values of $\beta$ and it is indicated that for the disk with 
$\beta \lesssim 0.1$, three-dimensional modes are more important than
 two-dimensional modes.

Since three-dimensional torque formula \eqref{tq_MR} is valid only
in the strong magnetic field limit, it is not possible to extrapolate 
this to the case with $\beta\sim 1$.  However, since the
torque formula for two-dimensional modes 
\eqref{tq_ART} does not have any restriction, we can safely conclude
that $k_z=0$ modes are always suppressed for strong magnetic field.
Therefore, qualitatively, we expect that two-dimensional modes are
suppressed for $\beta \lesssim 1$.  To verify this conjecture
quantitatively, we need careful analysis for $\beta \neq 0$ case, which
will be presented elsewhere (T. Muto and S. Inutsuka 2008, in
preparation).

\subsection{The Relation between Magnetic Field Strength and the
  Differential Torque}

For an unmagnetized disk, it is
known that the outer torque that is 
exerted by the disk outside the planet wins over the inner torque
exerted by the disk inside, when the disk gas density is larger in the
inner disk than the outer \cite{Ward97}.  This is the result of the
competition of two effects.  On one hand, since the inner density
is larger than the outer, the inner torque becomes larger than the
outer.  On the other hand, the effect called pressure buffer enhances
the outer torque.  Considering the background disk structure, gas is
slightly sub-Keplerian resulting from the outward pressure gradient.
Since the planet is in Keplerian rotation, the corotation point locates 
slightly inside the planet, which makes the outer Lindblad resonance
slightly closer to the planet.  Calculating the difference of these two
competing effects, the outer torque is larger than the inner torque.

Let us now qualitatively discuss the differential torque in the disk
with poloidal magnetic field.  First, we 
consider the two-dimensional mode when magnetic field exerted on the
disk inside the planet's orbit is larger than the outside.
In the case without variation in
density and temperature, the mild cutoff of the torque by magnetic
field makes the outer torque stronger.  If the disk has radially
decreasing magnetic pressure distribution, the planet locates slightly
outside the 
corotation point since outward magnetic pressure is exerted on the gas.
This also enhances the outer torque, since the outer effective Lindblad
resonance locates closer to the planet.  The outer torque is, therefore,
expected to be stronger because of these two effects, in contrast to the
toroidal magnetic field case of Terquem (2003) where the inner torque is
stronger.  This might indicate that the differential 
torque may be very sensitive to the configuration of magnetic field near
the planet.

\begin{figure}
 \plotone{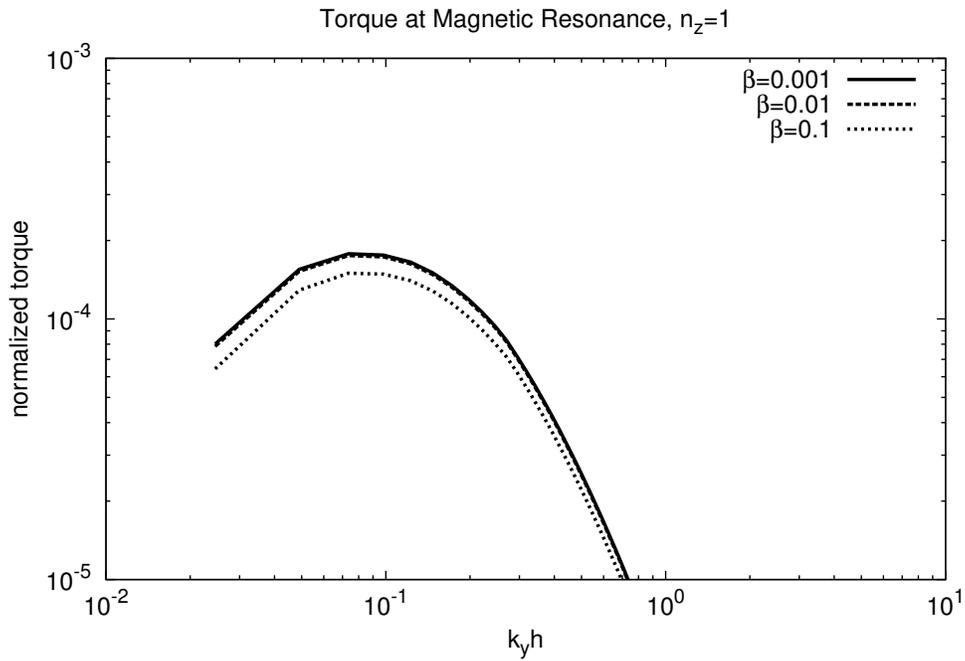}
 \figcaption{Torque at the magnetic resonance obtained by equation
 \eqref{tq_MR_mod}.  Magnetic field strength corresponds to
 $\beta=0.001$ (solid line), $\beta=0.01$ (dashed line), and $\beta=0.1$
 (dotted line).  The horizontal axis denotes azimuthal mode number
 and the vertical axis denotes normalized torque.
 \label{fig:mrtq_dis} }
\end{figure}

We now turn to the magnetic resonances of $k_z \neq 0$ modes, effective
for low $\beta$.  Since the formula \eqref{tq_MR} is for the limiting
case of $\beta \to 0$, the torque does not depend on the magnetic field 
strength.  We shall propose a simple torque formula for MR which
generalizes equation \eqref{tq_MR} and discuss the effect of
magnetic field.  Firstly, we note that the relation between $f$ and 
$\delta \rho / \rho_0$ is given by
\begin{equation}
 \dfrac{\delta\rho}{\rho_0} = \dfrac{1}{(c^2 + v_A^2) \sigma^2 - c^2
  v_A^2 k_z^2} \left[ \sigma^2 f - (\sigma^2 - v_A^2 k_z^2) \psi_p
	       \right].
\end{equation}
Since $f$ in the right hand side can be neglected when $\beta=0$, we
expect this term can be neglected even in $\beta\neq 0$, provided that
$\beta$ is sufficiently small.  This equation also indicates that there
is a $\delta$-function-like divergence at MR.  Neglecting the term with
$f$, we obtain the following torque formula at MR for low $\beta$,
\begin{equation}
  T_{\rm MR,mod} = \dfrac{2\pi}{3} L_y L_z \dfrac{ \rho_0 r_p
   k_z}{\Omega_p c (1+\beta)^{3/2}} \psi_{p,\rm MR}^2 ,
   \label{tq_MR_mod}
\end{equation}
where the value of gravitational potential is evaluated at MR.  The
smaller the magnetic field strength is, the closer towards the planet
the MR position locates, which makes the gravitational potential at MR,
$\psi_{\rm p, MR}$,
stronger. However, the coefficient, $(1+\beta)^{-3/2}$, becomes smaller,
which makes the evaluation complicated.  Using the parameters with
three-dimensional torque calculation, we evaluate the Fourier transform
of the gravitational potential and calculate the torque.  Figure
\ref{fig:mrtq_dis} shows the torque calculated from the modified formula
\eqref{tq_MR_mod} for $\beta=0.001$, $0.01$, and
$0.1$.  It is indicated that MR torque becomes smaller for weaker
magnetic field strength.  Therefore, we expect that when the inner
magnetic field is stronger than the outer magnetic field and the field
strength is high enough for $k_z\neq 0$ modes to be dominant, the inner
torque wins over outer torque, in analogous to  the results of the
analysis of toroidal field by Terquem (2003).

When magnetic field is very strong, equation
\eqref{tq_MR_mod} indicates that the value of the torque is not
sensitive to the strength of the field, and the differential torque can
be very small.  We consider, then, the effect of the gradient of
sound speed, which changes the location of the resonance even in
$\beta\to 0$ limit.  Let us consider the disk with higher sound speed
inside.
The outer MR, which nearly degenerates with VR, is closer to the planet,
giving a larger value of gravitational potential.  The
coefficient of the gravitational potential in equation \eqref{tq_MR}
is inversely proportional to the sound speed because $L_z k_z$ is the mode
number of the $z$-direction that is indifferent to the value of $c$.
Therefore, the coefficient is smaller for the inner MR than the outer MR.
The outer torque is expected to be stronger than the inner
torque when magnetic field is very strong and when there is a negative
gradient of the sound speed.

The qualitative dependence on magnetic field of the
differential torque for laminar modes may be summarized as follows.
Consider the case where inner magnetic field is stronger than the
outer.  When magnetic field strength is weak and two-dimensional modes
are dominant, the outer torque is more enhanced and the migration is
inward. When magnetic field is strong enough for three-dimensional
modes to be dominant, the migration can be outward.  Note, however, that
the rate and directions of migration may depend sensitively on the
gradient of the sound speed $c$. Negative gradient of the sound speed
may cause the inward migration for very low $\beta$.

It seems difficult to halt the inward migration in a disk with
strong poloidal magnetic field, since outward migration may require
positive gradient of sound speed.  Note, however, that the typical
magnitude of one-sided torque is 
always smaller than the unmagnetized case, 
as shown in figure \ref{fig:thin_2d}.  
In the disk with $\beta=100$, the torque is
dominated by two-dimensional modes and its magnitude is approximately
$10^{-3}$ in our normalization, while in $\beta=0.01$ case, the
magnitude is smaller by about two orders of magnitude.  
Therefore, the strong magnetization of disk is expected to 
slow down the migration. 
Actually this outcome is analogous to the effect of increasing 
gas temperature, and hence, thermal pressure and sound speed 
in the disk without magnetic field.

\subsection{Comparison with Toroidal Field Case}

In this paper, we have considered a protoplanetary disk threaded by 
 a poloidal magnetic field.  
In the analysis of the torque at the magnetic resonance, 
 we have considered a strong magnetic field case and
 derived a torque formula \eqref{tq_MR}.  
We now consider briefly the more general case when toroidal component 
 of magnetic field also exists.  
Although it is necessary to make a full, rigorous calculation 
 including both poloidal and toroidal components of magnetic field 
 in the background disk, we make a qualitative discussion by 
 comparing the effect of magnetic resonances of purely toroidal and 
 that of purely poloidal case. 

In the case of toroidal magnetic field, there is a magnetic
resonance in two-dimensional modes too (Terquem 2003).  In the strong
field limit, the position of the magnetic resonance is given by
\begin{equation}
 x_{\rm MR, toroidal} = \dfrac{2}{3} H .
\end{equation}
Since the magnetic resonance in poloidal case is located at
\begin{equation}
 x_{\rm MR, poloidal} = \dfrac{2}{3} \dfrac{k_z}{k_y} H ,
\end{equation}
and in a thin disk, modes
$k_z \gg k_y$ are important, the magnetic resonance of 
a toroidal field is closer to the planet than the toroidal case.  
We also note that the Fourier components of three-dimensional
modes of gravitational potential are much smaller than the
two-dimensional modes, as shown by Tanaka et al (2002).  
Therefore,
in a standard case of a planet on a circular orbit embedded in a disk
midplane, we expect that when the net magnetic field is dominated by
toroidal components, the effect of poloidal magnetic field is small
compared to that of a toroidal field.  
A possible exception is provided by a large inclination of 
 planet's orbit.
In this case, the gravitational potential
 of planet has a large z-component (three-dimensional modes), and thus, 
 the poloidal field would be important. 
Note, however, that we need a new set of analyses for the case of a
planet with inclined orbit (for unmagnetized disks,  see Tanaka \& Ward
2004).

\section{Summary} \label{summary}

We have performed linear perturbation analysis to calculate the torque
exerted on the planet embedded in the non-turbulent disk 
with poloidal magnetic field using the shearing sheet approximation.  We
have derived a second
order ordinary differential equation describing the excitation and
propagation of the wave, equation \eqref{eqn_prop}, and derived
analytic expressions of the torque for two limiting cases.  Equation
\eqref{tq_ART} gives the torque for two-dimensional modes.
Equation \eqref{tq_MR} gives the torque for three-dimensional modes for
$\beta \ll 1$ under WKB approximation.  We have
compared the result of the linear analysis and
numerical calculation, and found that both formulae show reasonable
agreement.  We have shown that the two-dimensional 
modes are suppressed when magnetic field is strong, in contrast
to Terquem (2003) analysis of toroidal field, indicating that the
property of planetary migration may be sensitive to the configuration of
the magnetic field around the planet.  It is also indicated that when
magnetic field is very strong, three-dimensional modes are more
effective than the two-dimensional modes, in contrast to the analysis of
the three-dimensional calculation of unmagnetized disk by Tanaka et
al. (2002) 

Since we have been using the shearing sheet approximation and
have derived the torque formulae only in restricted cases, the 
analysis of more general cases and other resonances is necessary, which
will appear elsewhere (T. Muto and S. Inutsuka 2008, in preparation).
Although the equation \eqref{tq_MR} agrees well with the
numerical calculations, this is derived under WKB approximation.
Therefore, we need careful analysis for high $k_y$ for three-dimensional
modes. We also need more quantitative analysis of the differential
torque, which determines the direction and rate of the migration of the 
planet.

\acknowledgments
The authors thank T. Matsumoto for the great help in carrying out
numerical calculations, and  F. Masset and C. Terquem for useful
discussions.
This work is supported by the Grant-in-Aid for the 21st Century COE
``Center for Diversity and Universality in Physics'' from the Ministry
of Education, Culture, Sports, Science and Technology (MEXT) of Japan.
Numerical calculations were in part carried
out on VPP5000 at the Center for Computational Astrophysics, CfCA, of
National Astronomical Observatory of Japan.
T. M. is supported by Grants-in-Aid for JSPS Fellows (19$\cdot$2409)
from MEXT of Japan.
M. M. is
supported by Grants-in-Aid (18740104) from MEXT
of Japan.
S. I. is
supported by Grants-in-Aid (15740118, 16077202, and 18540238) from MEXT
of Japan.

\appendix

\section{Derivation of Torque Formula for $k_z=0$ Modes}
\label{app:2dtq}

In this section, we derive the equation \eqref{tq_ART}, following
Artymowicz (1993) formalism.  The linearized equations of continuity and
motion for $k_z=0$ modes are
\begin{eqnarray}
& -i\sigma\dfrac{\delta \rho}{\rho_0} + \dfrac{d}{dx} \delta v_x + ik_y
 \delta v_y = 0, \\
& -i\sigma \delta v_x + c^2(1+\beta^{-1})\dfrac{d}{dx} \dfrac{\delta
 \rho}{\rho_0} - 2\Omega_p \delta v_y = - \dfrac{d}{dx} \psi_p, \\
& -i\sigma \delta v_y + \dfrac{1}{2} \Omega_p \delta v_x + ik_y
 c^2(1+\beta^{-1}) \dfrac{\delta \rho}{\rho_0} = -ik_y \psi_p.
\end{eqnarray}
From these, we obtain the equation for vorticity:
\begin{equation}
 \dfrac{d}{dx} \delta v_y - \dfrac{1}{2}\Omega_p \dfrac{\delta
  \rho}{\rho_0} - ik_y \delta v_x =0 .
\end{equation}
Using the equations of motion, we finally obtain the
 Schr\"{o}dinger-type second-order
ordinary differential equation for $\delta v_y$:
\begin{eqnarray}
 \dfrac{d^2}{dx^2} \delta v_y + \dfrac{1}{c^2 (1+\beta^{-1})} 
  \left[ \sigma^2 - \Omega_p^2  - c^2 (1+\beta^{-1})k_y^2  \right]  \delta
  v_y  \nonumber \\
 = \dfrac{1}{c^2(1+\beta^{-1})} \left[- \dfrac{1}{2}\Omega_p
  \dfrac{d \psi_p}{dx} + \sigma k_y \psi_p \right] .
  \label{paracyl_eqn}
\end{eqnarray}
The position of the effective Lindblad resonance is given by
\begin{equation}
 \sigma^2 = \Omega_p^2 + c^2(1+\beta^{-1})k_y^2 .
\end{equation}
Waves are evanescent in the region in the vicinity of the planet and the
regions further away from the resonance are propagation regions.  Waves
are excited at the effective Lindblad resonances and propagate away from
the planet to $|x| \to \infty$.

We shall calculate the angular
momentum flux at infinity.  The angular momentum flux is calculated by
\begin{equation}
 F_A = 2r_p \rho_0 L_y L_z \mathrm{Re} \left[ \delta v_x \delta
					v_y^{\ast} \right] .
\end{equation}
Since the gravitational potential of the planet vanishes at infinity,
the flux at the infinity is
\begin{equation}
 F_A(x\to\infty) = 2r_p \rho_0 L_y L_z \dfrac{4c^2(1+\beta^{-1})
  k_y}{\Omega_p^2 + c^2(1+\beta^{-1})k_y^2} \mathrm{Im} \left[ \delta
	 v_y^{\ast} \dfrac{d}{dx} \delta v_y \right] .
  \label{flux_inf}
\end{equation}
The solution of the wave equation \eqref{paracyl_eqn} is given by
parabolic cylinder functions.  Here, for simplicity, we assume the two
resonances, inside and outside the planet, are isolated each other.
The equation in the vicinity of the resonance is then given by
\begin{equation}
 \dfrac{d^2}{dz^2} \delta v_y + 2\gamma (z-\gamma) \delta v_y = -S,
\end{equation}
where
\begin{eqnarray}
& z = \left[ \dfrac{3\Omega_p k_y}{2c\sqrt{1+\beta^{-1}}}
     \right]^{\frac{1}{2}} x \\
& \gamma^2 = \dfrac{2}{3} \dfrac{\Omega_p^2+c^2(1+\beta^{-1})
 k_y^2}{\Omega_p ck_y \sqrt{1+\beta^{-1}}} \\
& S = \dfrac{\Omega_p}{c (1+\beta^{-1})^{3/4}\sqrt{6\Omega_p ck_y}}
 \left[ \dfrac{d\psi_p}{dz} - \sigma (1+\beta^{-1})^{1/4}
  \sqrt{\dfrac{8ck_y}{3\Omega_p^3}}\psi_p \right]_{\rm eff} .
\end{eqnarray}
The subscript ``eff'' denotes the quantity evaluated at the resonance.

We impose the boundary condition as follows.  In the evanescent region,
the solution does not grow exponentially and in the propagation region,
the waves propagate away from the planet.  The solution is then
\begin{equation}
 \delta v_y = \dfrac{\pi S}{(2\gamma)^{2/3}} \left\{
	 \mathrm{Gi}\left[-(2\gamma)^{1/3}(z-\gamma)\right] + i
	 \mathrm{Ai}\left[-(2\gamma)^{1/3}(z-\gamma)\right]
					     \right\}
 \label{sol_airy}
\end{equation}
Where $\mathrm{Ai}$ represents the Airy function and $\mathrm{Gi}$ is
the solution of the equation \\ \cite{ASfunc} 
\begin{equation}
 \dfrac{d^2}{dx^2}\mathrm{Gi}(x) - x \mathrm{Gi}(x) = -\dfrac{1}{\pi} .
\end{equation}
Substituting the equation \eqref{sol_airy} into \eqref{flux_inf}, we
obtain the torque formula \eqref{tq_ART}.

The condition of the isolation of the resonances is satisfied when the
solution at one 
resonance does not affect the other resonance.  The distance between the
resonances is $\delta z_{\rm res} \sim \gamma$, while the scale that the
solution in the vicinity of one resonance changes in the evanescent
region is $\delta z_{\rm wave} \sim\gamma^{-1/3}$.  Hence, resonances
are well separated each other when $\delta z_{\rm res}\gg \delta z_{\rm
wave}$, or $\gamma \gg 1$.

\section{The Evaluation of the Magnitude of $f$} \label{app:realf}

In this section, we evaluate the integral in the equation
\eqref{realf}.  For simplicity, we set the planet's gravitational
potential $\psi_p$ to be constant.
The part of the equation \eqref{realf},
\begin{equation}
 \int_{-\infty}^x du e^{k_z (u-x)} - \int_{+\infty}^x du e^{k_z(x-u)}
\end{equation}
is finite.  We consider the rest.  Since we are working on the local
Cartesian coordinate where inside and outside the planet are symmetric,
we assume $x>0$ without loss of generality.  Let $I$ be
\begin{eqnarray}
 I = \mathcal{P}\int_{-\infty}^{x} du e^{k_z(u-x)} \left[
						    \dfrac{1}{\sigma(u)-ck_z} 
			      - \dfrac{1}{\sigma(u)+ck_z}
					\right] \nonumber \\
 -  \mathcal{P}\int_{-\infty}^{x} du e^{k_z(x-u)} \left[
						   \dfrac{1}{\sigma(u)-ck_z} 
			      - \dfrac{1}{\sigma(u)+ck_z}
				       \right] ,
\end{eqnarray}
then
\begin{eqnarray}
& I = I_A - I_B - I_C + I_D , \\
& I_A = \exp\left[ -k_z\left( x - \dfrac{2ck_z}{3\Omega_p k_y} \right)
	    \right] \mathrm{Ei}\left( k_zx - \dfrac{2ck_z^2}{3\Omega_p
	    k_y} \right) , \\
& I_B = \exp\left[ -k_z\left( x + \dfrac{2ck_z}{3\Omega_p k_y} \right)
	    \right] \mathrm{Ei}\left( k_zx + \dfrac{2ck_z^2}{3\Omega_p
	    k_y} \right) , \\
& I_C = \exp\left[ k_z\left( x - \dfrac{2ck_z}{3\Omega_p k_y} \right)
	    \right] \mathrm{Ei}\left( - k_zx + \dfrac{2ck_z^2}{3\Omega_p
	    k_y} \right) , \\
& I_D = \exp\left[ k_z\left( x + \dfrac{2ck_z}{3\Omega_p k_y} \right)
	    \right] \mathrm{Ei}\left( - k_zx - \dfrac{2ck_z^2}{3\Omega_p
	    k_y} \right) , \\
\end{eqnarray}
where $\mathrm{Ei}$ denotes the exponential integral,
\begin{equation}
 \mathrm{Ei}(w) = \mathcal{P} \int_{-\infty}^w dt \dfrac{e^t}{t} .
\end{equation}

We can check $I_A$, $I_B$, $I_C$, and $I_D$ do not diverge at infinity
by the asymptotic expansion of the exponential integral,
\begin{equation}
 e^{-w} \int_{-\infty}^{w} \dfrac{e^t}{t}dt \sim \dfrac{1}{w}
\end{equation}
and the inequality derived from the series expansion of the exponential
integral,
\begin{equation}
 \mathrm{Ei}(w) = \gamma + \ln w + \sum_{n=1}^{\infty} \dfrac{w^n}{nn!}
  < \gamma  + \ln w + e^w .
\end{equation}
For $x \to \infty$, $I\sim \mathcal{O}(1)$.

We now consider the vicinity of the resonance, 
$x \sim 2ck_z/3\Omega_p k_y$. Although $I_A$ and $I_C$ are divergent
logarithmically, the combination $I_A - I_C$ does not diverge:
\begin{equation}
 I_A - I_C \sim \mathcal{P} \int_{-k_z \delta}^{k_z \delta} dt \dfrac{e^t}{t}
\end{equation}
where we set $x \sim 2ck_z/3\Omega_p k_y + \delta$.  
It is easy to show $I_B, I_D \sim \mathcal{O} (1)$ and therefore, $I$ is
order of unity for all $x$.  Although $\psi_p$ 
is not strictly a constant, it is smooth in the vicinity
of the resonance.  Therefore, we can safely assume this to be constant
when we discuss the divergence at the resonance.

In summary, from \eqref{realf}, the order of magnitude of
 $\mathrm{Re} f$ is 
\begin{equation}
 \mathrm{Re} f \sim \mathcal{O} \left( \dfrac{ck_z^2}{\Omega_p k_y} \psi_p
			\right) .
\end{equation}
It is clear from the equation \eqref{imf}, the imaginary part of $f$ is
also of the same order.

\clearpage

\clearpage

\clearpage

\clearpage

\clearpage

\clearpage

\clearpage

\clearpage

\clearpage

\clearpage

\end{document}